\documentclass[sigconf]{acmart}
\usepackage{amsmath,amssymb,amsfonts}
\usepackage[linesnumbered,ruled,vlined]{algorithm2e}
\usepackage{algorithmic}

\usepackage{balance}

\usepackage{calc}
\usepackage{colortbl}

\usepackage{enumitem}

\usepackage[T1]{fontenc}
\usepackage{float}

\usepackage{geometry}
\usepackage{graphicx}

\usepackage{listings}

\usepackage{makecell}
\usepackage{microtype}
\usepackage{multirow}

\usepackage{pifont}

\usepackage{rotating}

\usepackage{setspace}
\usepackage{soul}
\usepackage{subfigure}
\usepackage{stfloats}

\usepackage{textcomp}

\usepackage{url}

\usepackage{verbatim}

\usepackage{xcolor}
\usepackage{xspace}
\newcommand{\parabf}[1]{\noindent\textbf{#1}}

\definecolor{ggray}{HTML}{eff0f0}
\definecolor{gggray}{HTML}{E8E8E8}
\definecolor{ggggray}{HTML}{BEBEBE}
\definecolor{myyellow}{HTML}{FFF2CC}

\newcommand{\app}{KTester\xspace}

\newcommand{\ie}{\textit{i.e.,}\xspace}
\newcommand{\eg}{\textit{e.g.,}\xspace}

\newcounter{finding}
\newcommand{\finding}[1]{
  \refstepcounter{finding}
  \vspace{0.2\baselineskip}
  \noindent
  \colorbox{myyellow!30}{
    \parbox{0.96\linewidth}{
      \vspace{1pt}
      \textbf{Finding~\arabic{finding}:} #1
      \vspace{1pt}
    }
  }
}

\AtBeginDocument{%
  }

\copyrightyear{2026}
\acmYear{2026}
\setcopyright{cc}
\setcctype{by-nc-nd}
\acmConference[ICSE '26]{2026 IEEE/ACM 48th International Conference on Software Engineering}{April 12--18, 2026}{Rio de Janeiro, Brazil}
\acmBooktitle{2026 IEEE/ACM 48th International Conference on Software Engineering (ICSE '26), April 12--18, 2026, Rio de Janeiro, Brazil}
\acmPrice{}
\acmDOI{10.1145/3744916.3787769}
\acmISBN{979-8-4007-2025-3/2026/04}

\begin{document}

\title{Knowledge Matters: Injecting Project and Testing Knowledge into LLM-based Unit Test Generation}

%%
%% The "author" command and its associated commands are used to define
%% the authors and their affiliations.
%% Of note is the shared affiliation of the first two authors, and the
%% "authornote" and "authornotemark" commands
%% used to denote shared contribution to the research.

\author{Anji Li}
% \email{lianj8@mail2.sysu.edu.cn}
\orcid{0009-0001-4881-4148}
\affiliation{%
  % \department{Key Laboratory of Trusted Large Language Models}
  \institution{Sun Yat-sen University}
  \city{Zhuhai}
  \country{China}}

\author{Mingwei Liu*}
% \email{liumw26@mail.sysu.edu.cn}
\orcid{0000-0002-3462-997X}
\affiliation{%
  % \department{Key Laboratory of Trusted Large Language Models}
  \institution{Sun Yat-sen University}
  \city{Zhuhai}
  \country{China}}

\author{Zhenxi Chen}
% \email{chenzhx236@mail2.sysu.edu.cn}
\orcid{0009-0000-0896-3811}
\affiliation{%
  % \department{Key Laboratory of Trusted Large Language Models}
  \institution{Sun Yat-sen University}
  \city{Zhuhai}
  \country{China}}

\author{Zheng Pei}
% \email{peizh3@mail2.sysu.edu.cn}
\orcid{0009-0001-5127-7750}
\affiliation{%
  % \department{Key Laboratory of Trusted Large Language Models}
  \institution{Sun Yat-sen University}
  \city{Zhuhai}
  \country{China}}

\author{Zike Li}
\orcid{0009-0000-0807-1805}
% \email{lizk8@mail2.sysu.edu.cn}
\affiliation{%
  % \department{Key Laboratory of Trusted Large Language Models}
  \institution{Sun Yat-sen University}
  \city{Zhuhai}
  \country{China}}

\author{Dekun Dai}
% \email{daidk@mail2.sysu.edu.cn}
\orcid{0009-0002-1452-8562}
\affiliation{%
  % \department{Key Laboratory of Trusted Large Language Models}
  \institution{Sun Yat-sen University}
  \city{Zhuhai}
  \country{China}}

\author{Yanlin Wang}
% \email{wangylin36@mail.sysu.edu.cn}
\orcid{0000-0001-7761-7269}
\affiliation{%
  % \department{Key Laboratory of Trusted Large Language Models}
  \institution{Sun Yat-sen University}
  \city{Zhuhai}
  \country{China}}

\author{Zibin Zheng}
\orcid{0000-0002-7878-4330}
% \email{zhzibin@mail.sysu.edu.cn}
\affiliation{%
  % \department{Key Laboratory of Trusted Large Language Models}
  \institution{Sun Yat-sen University}
  \city{Zhuhai}
  \country{China}}

\thanks{
*M. Liu is the corresponding author.

A. Li, M. Liu, Z. Chen, P. Zheng, Z. Li, D. Dai, Y. Wang and Z. Zheng are with the School of Software Engineering and the Zhuhai Key Laboratory of Trusted Large Language Models, Sun Yat-sen University, Zhuhai, China.

Email: lianj8@mail2.sysu.edu.cn, liumw26@mail.sysu.edu.cn, \{chenzhx236, peizh3, lizk8,daidk\}@mail2.sysu.edu.cn, \{wangylin36, zhzibin\}@mail.sysu.edu.cn.
}

%%
%% By default, the full list of authors will be used in the page
%% headers. Often, this list is too long, and will overlap
%% other information printed in the page headers. This command allows
%% the author to define a more concise list
%% of authors' names for this purpose.
% \renewcommand{\shortauthors}{Anji Li et al.}

\begin{abstract}
    Automated unit test generation using large language models (LLMs) holds great promise but often struggles with generating tests that are both correct and maintainable in real-world projects. This paper presents \app, a novel framework that integrates project-specific knowledge and testing domain knowledge to enhance LLM-based test generation. Our approach first extracts project structure and usage knowledge through static analysis, which provides rich context for the model. It then employs a testing-domain-knowledge-guided separation of test case design and test method generation, combined with a multi-perspective prompting strategy that guides the LLM to consider diverse testing heuristics. The generated tests follow structured templates, improving clarity and maintainability. We evaluate \app on multiple open-source projects, comparing it against state-of-the-art LLM-based baselines using automatic correctness and coverage metrics, as well as a human study assessing readability and maintainability. Results demonstrate that \app significantly outperforms existing methods across six key metrics, improving execution pass rate by 5.03\% and line coverage by 11.67\% over the strongest baseline, while requiring less time and generating fewer test cases. Human evaluators also rate the tests produced by \app significantly higher in terms of correctness, readability, and maintainability, confirming the practical advantages of our knowledge-driven framework.
\end{abstract}

%%
%% The code below is generated by the tool at http://dl.acm.org/ccs.cfm.
%% Please copy and paste the code instead of the example below.
%%


% \keywords{Do, Not, Us, This, Code, Put, the, Correct, Terms, for,
  % Your, Paper}

\maketitle

\section{Introduction}
\label{sec:intro}
Unit testing is a fundamental practice in software development that verifies whether a method behaves as expected, playing a key role in ensuring software correctness, maintainability, and enabling regression testing~\cite{leung1989insights}. As the first line of defense in the development lifecycle, it helps detect and localize bugs early, preventing their propagation~\cite{runeson2006survey}. However, writing high-quality unit tests manually is often time-consuming and error-prone, as it requires developers to deeply understand the method’s behavior, construct valid input states, and define precise assertions~\cite{daka2014survey}.

For a method under test (\ie{} often called the focal method), a well-constructed unit test usually has two parts: the test prefix, which prepares the objects and environment needed to bring the system into a testable state, and the test oracle, which verifies that the actual output matches the expected behavior~\cite{DBLP:journals/tse/BarrHMSY15}. High-quality unit tests improve software reliability and also serve as living documentation, facilitating code comprehension and maintenance~\cite{fraser2014large}.

\looseness=-1
To reduce manual effort, automated test generation techniques have been proposed to generate a suite of unit tests with the main goal of maximizing the coverage in the software under test. Traditional unit test generation techniques include search-based~\cite{harman2009theoretical, blasi2022call, delgado2022interevo}, constraint-based methods~\cite{ernst2007daikon, csallner2008dysy, xiao2013characteristic}, and random-based strategies~\cite{zeller2019fuzzing, pacheco2007feedback}. While effective in achieving code coverage, the generated tests are often hard to read and maintain, limiting practical adoption~\cite{almasi2017industrial}.

\begin{figure}[thbp]
    \centering
    \includegraphics[width=0.98\columnwidth]{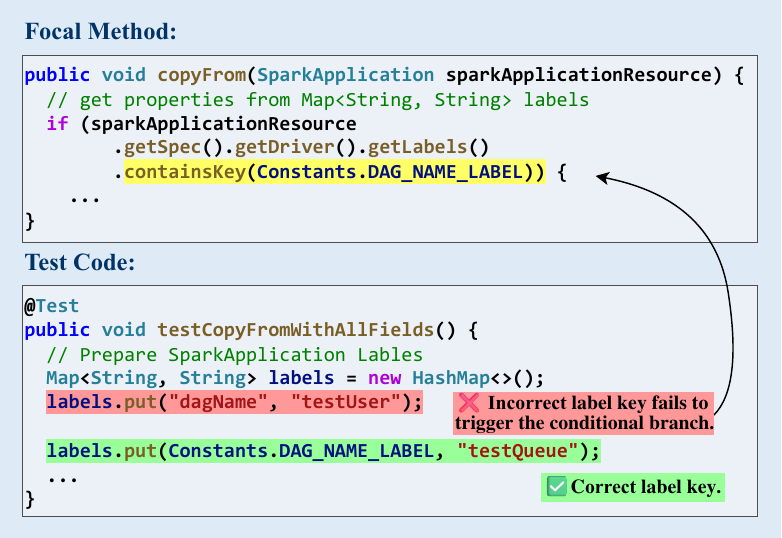}
    \caption{Motivational Examples(a): Setting Tested Object Incorrectly}
    \Description[Motivational Examples A]{Motivational Examples(a): Setting Tested Object Incorrectly}
    \label{fig:motivation-a}
\end{figure}

Recently, large language models (LLMs) have shown great promise in generating more human-like test code. These LLM-based methods demonstrate promising capabilities in understanding code semantics and synthesizing test code without explicit test specifications. Despite this potential, current LLM-based methods still suffer from key limitations that hinder their practical adoption due to a fundamental lack of essential knowledge. In particular, these models often lack access to \textbf{project-specific knowledge}, such as how to correctly instantiate and use classes, or how utility methods and APIs interact across modules. They also overlook \textbf{testing domain knowledge}, including core principles like boundary value analysis, exception handling, or the separation between test design and implementation. \textbf{This lack of both project-specific and testing-domain knowledge often results in test code that is unreliable and difficult to maintain.}

\begin{figure}[t]
    \centering
    \includegraphics[width=0.95\columnwidth]{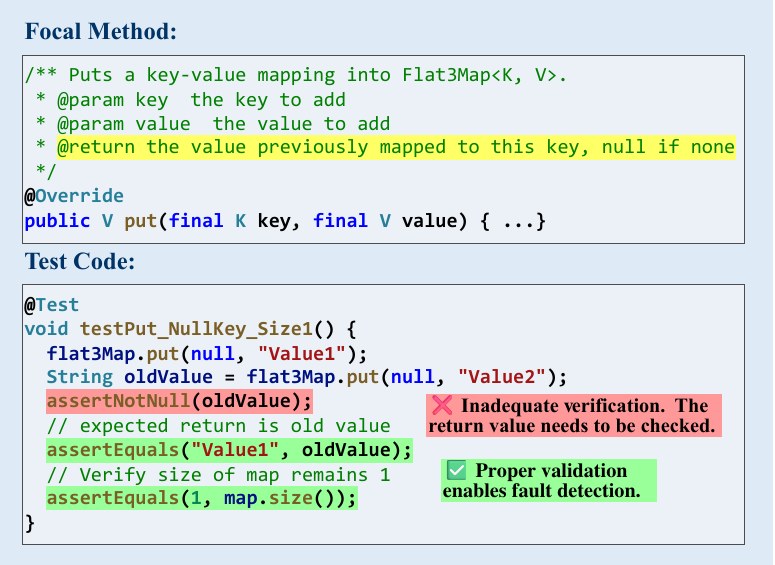}
    \caption{Motivational Examples(b): Insufficient Assert Statements}
    \Description[Motivation Examples B]{Motivational Examples(b): Insufficient Assert Statements}
    \label{fig:motivation-b}
\end{figure}

\looseness=-1
As illustrated in Figure~\ref{fig:motivation-a}-\ref{fig:motivation-c}, these issues commonly stem from the lack of project-specific knowledge and test domain expertise in existing LLM-based test generation methods. For instance, Figure~\ref{fig:motivation-a} shows an incorrect construction of the \textit{labels} Map used by \textit{SparkApplication}, inserting the key ``dragName'' instead of the correct constant \textit{Constants.DAG\_NAME\_LABE}. This error arises from a lack of awareness of constructor dependencies and configuration requirements. Figure~\ref{fig:motivation-b} depicts a test with insufficient assert statements, weakening its effectiveness. Meanwhile, Figure~\ref{fig:motivation-c} shows an inconsistent test structure with hard-coded values and missing essential setup logic, which results in fragile and difficult-to-maintain code.

To overcome limitations in existing LLM-based unit test generation, we propose a novel framework that integrates project-specific knowledge with software testing domain knowledge to guide the test generation process. Our approach consists of two main phases: an offline knowledge extraction phase and an online test generation pipeline. In the offline phase, we perform static analysis on the target project to extract two key types of knowledge: project structure knowledge, which includes class definitions, method signatures, and field declarations; and project usage knowledge, which covers invocation patterns, dependencies, and related functions. This comprehensive project knowledge is used to provide rich contextual information that improves the relevance and accuracy of the generated tests. The online test generation pipeline involves five sequential steps: (1) test class framework generation, which sets up reusable scaffolding such as setup and teardown methods; (2) multi-perspective test case design, where test scenarios are created from complementary testing heuristics; (3) test method transformation, which converts structured test cases into executable test methods; (4) test class integration, assembling all generated methods into a coherent test class; and (5) test class refinement, which improves code quality and maintainability. Throughout these steps, the pipeline leverages both the extracted project knowledge and guidance from established software testing principles.

Our approach’s core innovations are twofold. First, \textbf{the explicit incorporation of project-aware knowledge provides the LLM with rich, contextualized information about the codebase}, leading to more accurate and meaningful tests. Second, we introduce a \textbf{testing domain knowledge–guided separation of test design and test generation and multi-perspective test case design prompting strategy}. By decoupling the “what to test” from “how to test”, we make the testing intent explicit and improve the semantic clarity and maintainability of the generated test code.

We validate our approach on several real-world open-source projects, comparing it against state-of-the-art LLM-based baselines using both automated metrics (e.g., execution pass rate, line coverage) and a human study assessing readability, maintainability, and test intent clarity. Our findings show that (1) our method consistently outperforms baselines across six key metrics, improving execution pass rate by 5.03\% and line coverage by 11.67\% compared to the strongest baseline, while takes shorter time and generate less test cases; (2) the modular test case transformation component, which embodies the testing-principle-guided design, has the largest individual impact, with removal causing a drop of 11.45\% in execution pass rate and 13.39\% in line coverage; and (3) human evaluators rate our generated tests significantly higher in correctness, readability, and maintainability, confirming the practical advantages of our knowledge-driven framework.

Our main contributions are:

\begin{figure}[]
    \centering
    \includegraphics[width=0.9\columnwidth]{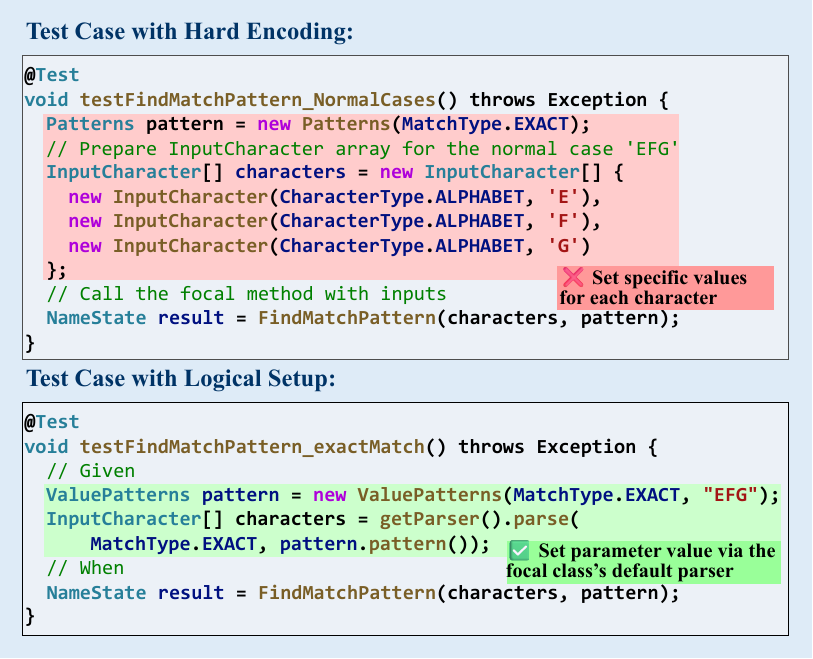}
    \caption{Motivational Examples(c): Hard Encoding Values}
    \label{fig:motivation-c}
    \Description[Motivational Examples(c)]{Motivational Examples(c): Hard Encoding Values}
\end{figure}

\begin{itemize}
    \item We propose a novel test generation framework that integrates project-specific knowledge with testing domain knowledge. By leveraging comprehensive project knowledge and applying a separation of test design and test code generation guided by testing domain knowledge, our approach produces more accurate and maintainable test cases.
     \item We introduce a multi-perspective prompting strategy along with testing-domain-knowledge-guided separation of test case design and test method implementation, enabling the LLM to generate semantically rich, logically structured, and purposeful tests.
    \item We perform extensive evaluation, including automatic metrics and human studies, demonstrating that our method consistently outperforms existing LLM-based approaches in correctness, readability, maintainability, and overall quality.
\end{itemize}

All data/code used in this study is provided in the package~\cite{Replication_Package}. 

%\section{Motivation} 
%\label{sec:motivation}
%\input{sections/motivation}

\section{Approach}
\label{sec:app}
We present \app, a knowledge-aware unit test generation framework that leverages both project-specific knowledge and testing domain knowledge to guide LLMs in generating high-quality, semantically meaningful, and maintainable unit tests.

Unlike prior LLM-based approaches that rely solely on the focal method as input and often generate brittle or unclear test cases, \app decouples test design from test code generation, and injects knowledge at multiple stages to address key deficiencies of prior work (see Figure~\ref{fig:motivation-a}, \ref{fig:motivation-b} and \ref{fig:motivation-c}). In particular, \app is designed around the following core principles:

\textbf{Project Knowledge Awareness.} We statically analyze the entire codebase to extract structured information about classes, methods, usage patterns, dependencies, and documentation. This knowledge is later used to resolve object instantiation, locate related APIs, and reduce hallucinations during code generation.

\begin{figure}[thbh]
    \centering
    \includegraphics[width=\linewidth]{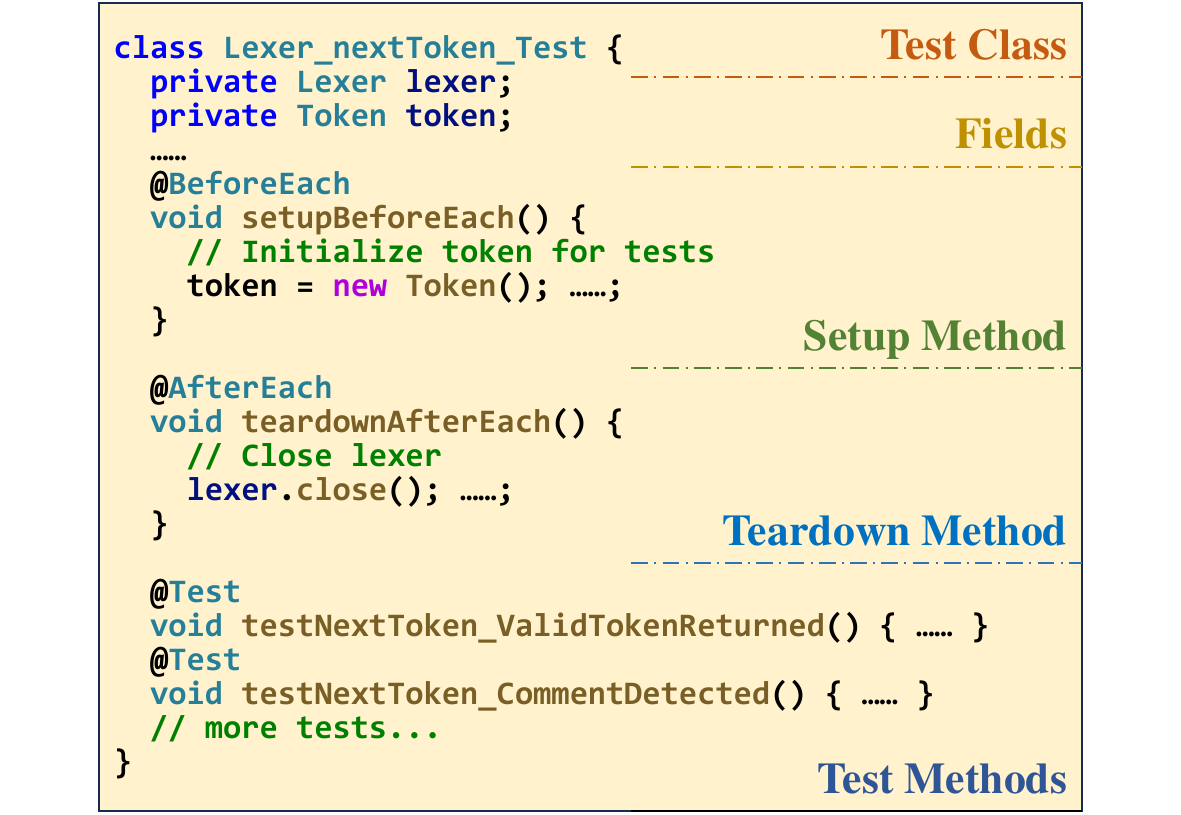}
    \caption{Test Class Example}
    \Description[Test Class Example]{The framework of test class, includes classes, methods, usage patterns, dependencies, and documentation.}
    \label{fig:test-class-example}
\end{figure}

\textbf{Testing Knowledge Awareness. We explicitly guide the LLM to design test cases from multiple testing perspectives (e.g., control flow, boundary, exception handling), instead of treating the task as a black-box code-to-code translation.}

\textbf{Modular Generation Pipeline. We separate test class framework construction, scenario-specific test design, and final test case synthesis into distinct stages, improving modularity, reusability, and interpretability of generated tests.}

\begin{figure*}
    \centering
    \includegraphics[width=0.9\textwidth]{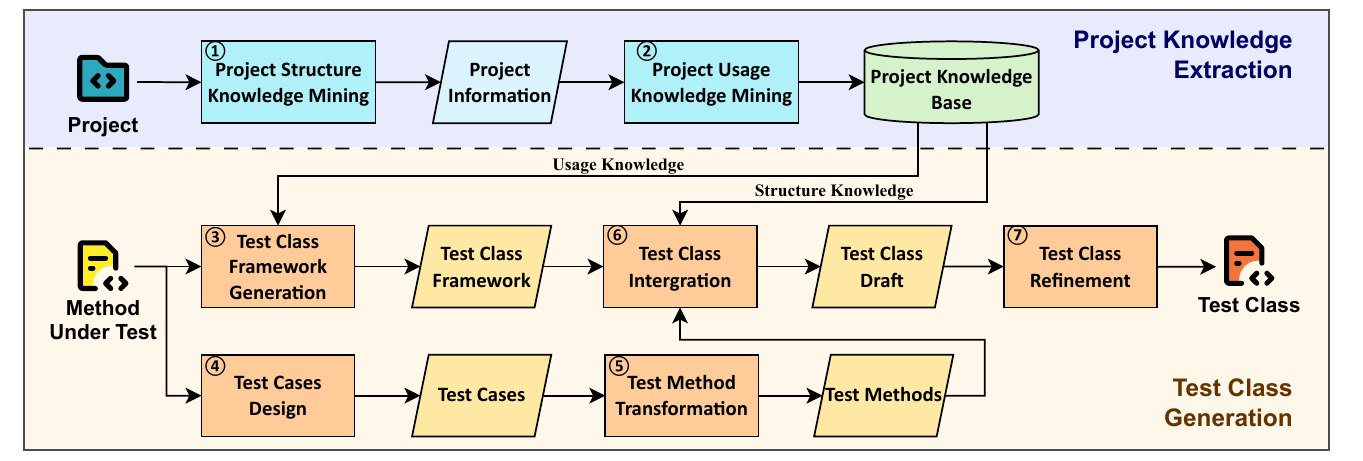}
    \caption{The Framework of \app}
    \Description[The Framework of KTester]{The Framework of KTester}
    \label{fig:framework}
\end{figure*}

\looseness=-1
The two-stages process of KTester generating the test class is showed in Figure \ref{fig:framework}. First, during the offline knowledge extraction stage (Section~\ref{sec:knowledge_extraction}), we perform static analysis of the target project to build a structured knowledge base capturing essential project-specific information such as class hierarchies, method signatures, field access patterns, call relationships, and document comments. This knowledge base provides a foundation for accurate, context-aware test generation. Second, in the online test generation stage (Section~\ref{sec:test_generation}), given a focal method, \app retrieves relevant project knowledge and testing heuristics to construct a knowledge-rich prompt for the LLM. The LLM then produces a structured test class framework along with diverse test cases, which are validated and optionally repaired to ensure correctness and completeness. Figure~\ref{fig:test-class-example} illustrates an example generated test class, including field declarations, setup and teardown methods, and multiple test methods.

\subsection{Project Knowledge Extraction}
\label{sec:knowledge_extraction}

\looseness=-1
To mitigate the limitations of LLMs in understanding project-specific semantics, \app performs a comprehensive offline analysis to extract structured \textit{project knowledge} from the codebase. This process is based on the assumption that unit tests are to be generated for methods within a project, and thus, knowledge beyond the focal method—such as class definitions, field usage, method invocations, and usage examples—is essential for meaningful test generation.

Specifically, \app conducts two key analyses to construct a reusable project knowledge base:
project structure knowledge mining, which captures the static architecture of the codebase—including class hierarchies, method signatures, and inter-method dependencies—and
project usage knowledge mining, which extracts realistic invocation contexts that reveal how focal methods are constructed, initialized, and invoked in real code scenarios.

\subsubsection{Project Structure Knowledge Mining}
To enable knowledge-aware unit test generation, \app statically analyzes the source code to mine structured \textit{project structure knowledge}. This includes the architectural, semantic, and dependency-level information necessary for understanding the focal method’s context and generating correct, maintainable test code. The analysis is based on abstract syntax tree (AST) parsing and code graph construction, allowing us to systematically extract and index key properties of classes and methods across the project.

Specifically, for each class and its methods or constructors, we extract the following information:

\parabf{Structural Metadata.} 
\looseness=-1
We collect the class name, package path, field declarations, constructor signatures, and method signatures (including parameter types and return types). This information enables the model to understand how to instantiate the target class or construct inputs for the method under test. For example, if a method requires a parameter of type \texttt{UserConfig}, this metadata reveals how \texttt{UserConfig} can be constructed via its fields and constructors.

\parabf{Document Comments.} We extract Javadoc-style comments associated with classes, constructors, and methods, as they often contain high-level semantic descriptions of behavior and usage constraints. These comments help LLMs infer testing intent and expected behavior. For instance, a comment like \texttt{/* Returns null if no user is found */} suggests that null should be included as a valid return value in test oracles.

\parabf{Dependency Relations.} For each method or constructor, we extract the list of invoked methods and accessed fields. These relationships are fundamental for understanding how the method interacts with its internal state or other classes. For example, if method \texttt{updateBalance()} internally calls \texttt{validateAccount()} and accesses \texttt{this.balance}, such dependencies indicate that tests should ensure the account is in a valid state and that balance updates are properly asserted. These dependency relations also support the identification of functionally related methods (used later in Usage Context Construction) and enable accurate call path tracing for generating invocation examples (Section~\ref{sec:usage_context}).

\subsubsection{Project Usage Knowledge Mining}
\label{sec:usage_context}
To mitigate the limitations of LLMs in understanding project-specific usage patterns, we extract \textit{Project Usage Knowledge} from the source code. This knowledge captures realistic invocation scenarios of the focal method, including (1) how its input parameters are constructed and (2) how the target object is initialized—both of which are essential for generating executable and semantically meaningful unit tests.

\begin{figure}
   \centering
   \includegraphics[width=\linewidth]{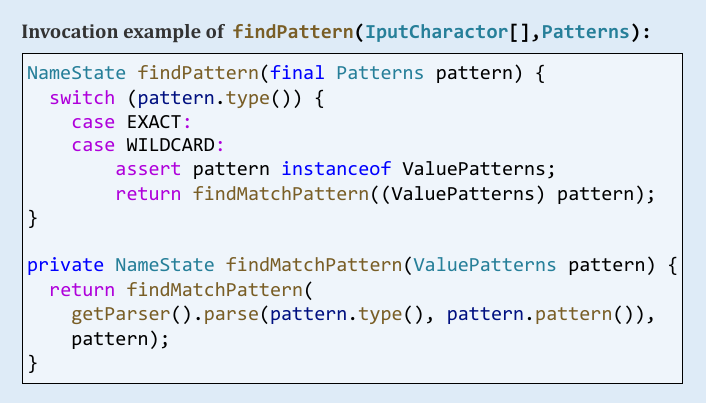}
   \caption{Illustration of extracting invocation examples from a caller method via control flow pruning.}
   \Description[Invocation Examples]{Illustration of extracting invocation examples from a caller method via control flow pruning.}
   \label{fig:invocation_example}
\end{figure}

Such usage knowledge provides concrete, in-project examples that reflect actual calling contexts, enabling the LLM to synthesize inputs that align with real usage scenarios. For example, it reveals how the focal class is typically constructed in the project, which may involve complex initialization logic or dependency injection. It also uncovers how input parameters are derived—such as being composed from helper methods, shared resources, or intermediate computations—rather than being hardcoded literals. By grounding the generation process in real-world usage, this approach improves both the realism and maintainability of the generated test cases. As the code snippet illustrated in Figure~\ref{fig:invocation_example}, the focal method \textit{findMatchPattern(InputCharactor[], Patterns)} is invoked inside \textit{findMatchPattern(ValuePatterns)} after checking whether the variable \textit{pattern} is an instance of \textit{ValuePatterns} class, and called a Parser to parse this pattern into an \textit{InputCharactor} array. Such code would be extracted as representative usage knowledge.

To mine project usage knowledge, we first identify the relevant caller context and then extract semantically meaningful execution traces showing how the focal method is invoked in practice.

\parabf{Caller Method Discovery.} For each focal method, we identify its \textit{caller methods} using the static call graph built during the offline analysis. When the focal method is private or not directly reachable from public APIs, we trace the shortest call chain from an externally accessible method to the focal method. This ensures that the usage knowledge reflects a realistic and complete setup process, capturing how the focal method is constructed and triggered in actual usage.

\parabf{Path-sensitive Usage Trace Extraction.} For each discovered caller method, we construct a Control Flow Graph (CFG)~\cite{allen1970control} and perform intra-procedural slicing to extract execution paths from the method’s entry point to the focal method invocation. By pruning unrelated branches and removing irrelevant operations, we isolate minimal yet semantically rich usage traces. These traces, which demonstrate how the focal class is instantiated and how arguments are prepared, are incorporated into the generation prompt as part of the Project Usage Knowledge. This provides the LLM with grounded, context-aware examples that help it synthesize test inputs aligned with the project’s actual usage conventions.

\subsubsection{Function-Level Index Construction}
\looseness=-1
To enable scalable and consistent test generation, each method or constructor is transformed into a \textit{knowledge unit} with its signature, documentation, and dependencies (invoked methods and accessed fields). These units are stored in an indexable format, forming a function-level knowledge base that enables efficient retrieval of related methods, usage contexts, and class-level information during prompt construction.

Unlike ad hoc, per-method extraction, our analysis is conducted \textit{once per project} and reused across focal methods. The knowledge base is also \textit{incrementally updatable}, requiring reprocessing only changed entities, which makes it practical for evolving codebases. While our method is language-agnostic, we focus on Java due to its popularity and mature analysis tooling\cite{javaparser,spoon_2025}. We use Spoon~\cite{spoon_2025} to efficiently extract ASTs, control flow, and dependency graphs.
\subsection{Test Class Generation}
\label{sec:test_generation}

As illustrated in Figure~\ref{fig:framework}, \app generates a complete and executable test class for a given focal method by leveraging offline-extracted project knowledge and domain-specific testing expertise. It constructs a rich generation context that integrates structural program information, usage patterns, and unit testing heuristics, guiding the LLM to produce high-quality, realistic unit tests. The generation pipeline consists of five steps: test class framework generation, test case design, test method transformation, test class integration, and test class refinement.

Inspired by how developers and testers write unit tests in practice, \app mimics the typical testing workflow—first planning test cases based on expected behaviors and coverage goals, then gradually transforming them into executable test code—rather than relying on a single-step code generation process.

\begin{figure}
    \centering
    \includegraphics[width=0.9\linewidth]{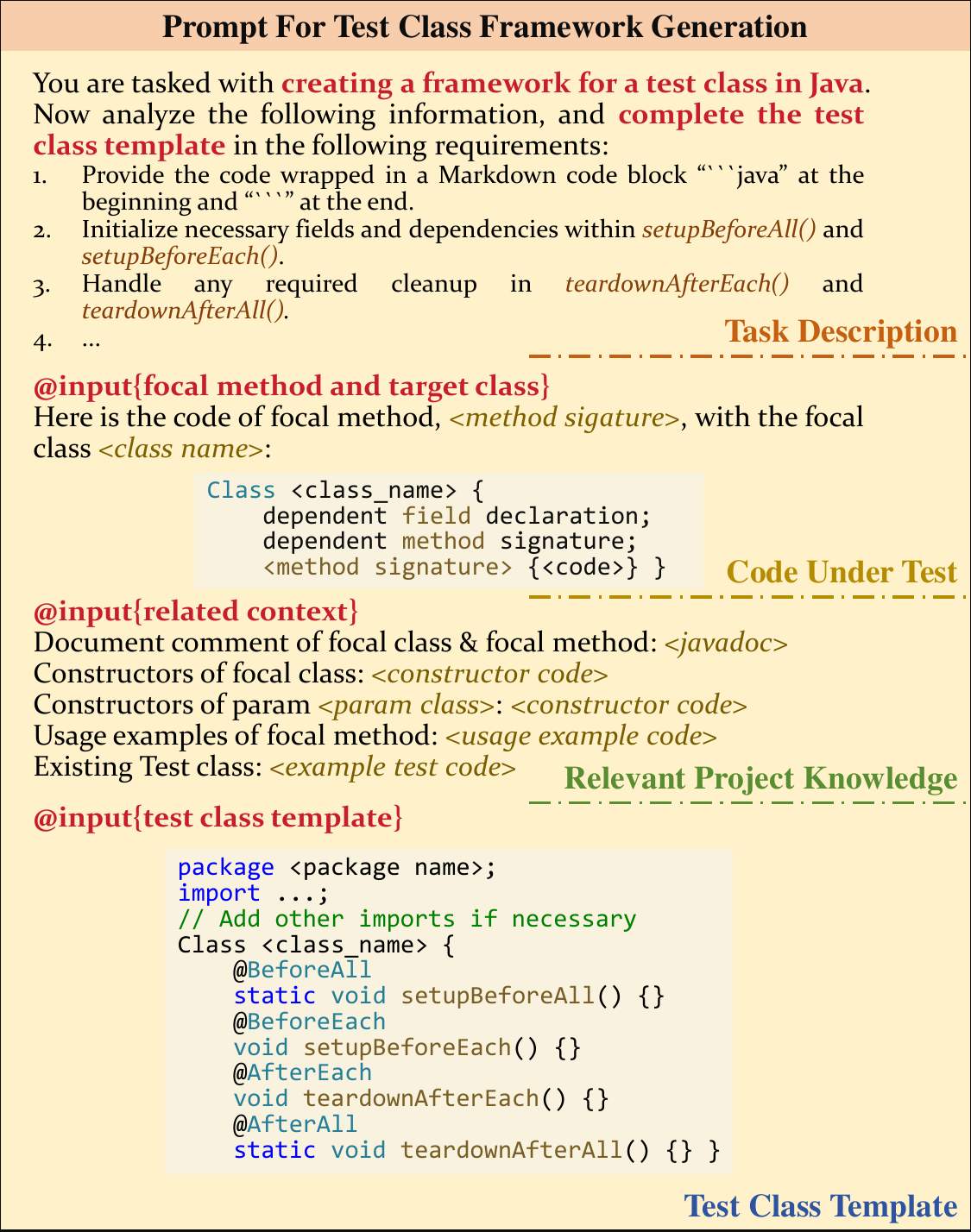}
    \caption{Prompt for test class framework generation.}
    \Description[Prompt for test class framework generation.]{Prompt for test class framework generation.}
    \label{fig:prompt-frame-gen}
\end{figure}

\subsubsection{Test Class Framework Generation}
\label{sec:app:framework-gen}
The first step prompts the LLM to generate a test class framework that establishes the necessary environment for unit testing.
Unlike prior methods~\cite{chen_chatunitest_2024,wang_hits_2024} that directly prompt the LLM with a focal method and expect complete test cases in one shot, \app decouples the generation process by first constructing a reusable, project-aware test class framework. This decision is grounded in a key insight: test generation is not merely a method-level translation task, but a context-sensitive activity requiring awareness of surrounding project structure, usage conventions, and domain-specific testing patterns. As illustrated in Figure~\ref{fig:prompt-frame-gen}, we design a prompt template grounded in project knowledge and test domain knowledge to guide this process. The prompt is composed of several key components, including core task and instructions, focal method and class, relevant project knowledge, test class template, and output specification.

The \textbf{core task and instructions} section provides high-level guidance to the LLM on setting up proper initialization and cleanup routines, enforcing access restrictions to target class internals, and ensuring syntactic correctness, thereby enabling a robust and maintainable test framework. The \textbf{focal method and class} section specifies the method under test and its containing class, providing the primary scope for framework generation. The \textbf{output specification} section defines the expected format and content of the generated framework, ensuring consistency and completeness.

To enrich the context, the \textbf{relevant project knowledge} section of the prompt draws from an offline-constructed knowledge base that captures both semantic and structural details of the codebase. This includes document comments for the focal class and method, which describe functionality, preconditions, and side effects to inform meaningful setup and teardown procedures. It also comprises constructors and parameter details, assisting the LLM in object instantiation and input preparation, particularly in complex scenarios. Additionally, usage knowledge mined from the codebase provides real-world usage patterns of the focal method and related classes, grounding the test framework in practical contexts. Finally, existing test classes associated with the focal class are incorporated as references for mocking strategies, resource management, and testing conventions, with only setup logic and class-level configurations extracted to avoid redundancy.

The \textbf{test class template} section outlines the structural skeleton, including annotations, lifecycle methods (e.g., setup and teardown), and placeholders for field declarations. This template, informed by practical experience in writing test code, serves as a starting point for the LLM to fill in with project-specific logic and configurations.

By integrating these components, \app creates context-aware prompts that enable the LLM to generate reusable, well-structured test class frameworks aligned with project conventions. This modular approach enhances consistency and reduces errors common in one-shot generation. Separating framework from test logic also enables reuse across methods, minimizing redundant effort and reflecting practical testing workflows.

\begin{figure}[t]
    \centering
    \includegraphics[width=0.9\columnwidth]{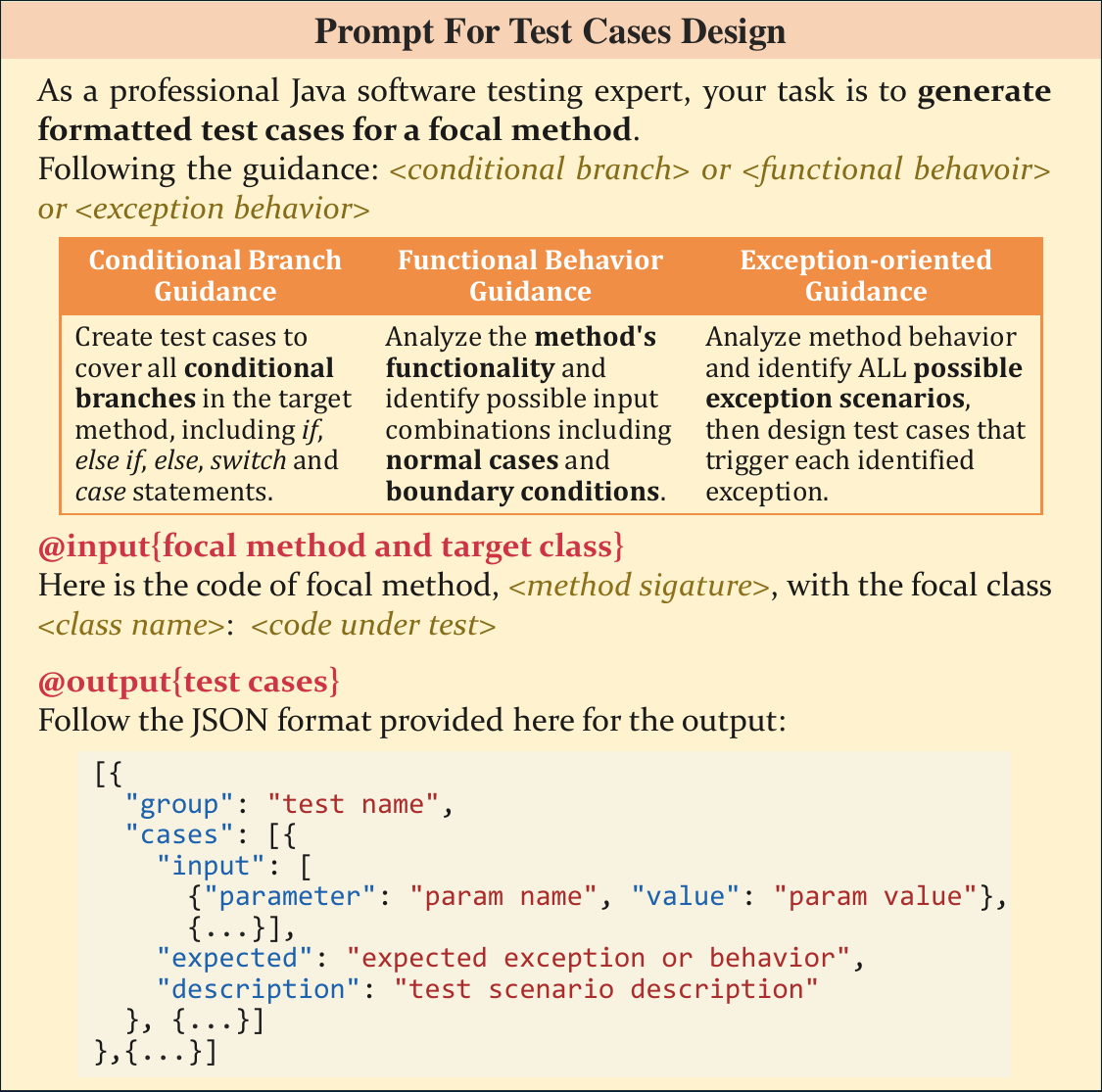}
    \caption{Prompts for multi-view test case design.}
    \Description[Prompts for multi-view test case design.]{Prompts for multi-view test case design.}
    \label{fig:prompt-test-case-design}
\end{figure}

\subsubsection{Test Cases Design}
\label{sec:test-case-design}
After generating the test class framework, the next step is to design high-quality test cases that cover different functional and non-functional behaviors of the focal method. In this phase, \app focuses on the planning of test cases, rather than directly generating test code. Unlike the previous step, test case design solely leverages domain-specific testing knowledge, without relying on project-specific context. The only input is the focal method itself.

\looseness=-1
\app adopts a multi-view guidance strategy grounded in established software testing principles. The LLM is explicitly instructed to design test cases from multiple perspectives—such as functional behavior, boundary conditions, and exception handling—and to organize them into logical groups based on shared testing intent. Each test case is represented internally in a machine-readable intermediate format that records structured natural-language descriptions of the scenario. In our implementation, this intermediate representation is realized using a lightweight JSON structure, and an illustrative example is provided in our replication package~\cite{Replication_Package}. This design intentionally decouples test intent from concrete executable code, improving clarity and maintainability. The intermediate representations are subsequently translated into framework-specific test code.

To facilitate structured generation, the LLM is required to output grouped test cases, where each group targets a distinct aspect of the focal method. For example, one group may cover valid inputs for core functionality, another may focus on boundary values such as empty lists or null, and a third may address failure or exception-triggering scenarios. This grouped organization improves modularity and controllability during subsequent transformation steps and mirrors human testing practices, where related scenarios are often clustered for clarity and reuse.

To guide the LLM, we design three types of prompts, each corresponding to a classical testing perspective, as shown in Figure~\ref{fig:prompt-test-case-design}. These prompts share a common structure, with variations only in the design guidance section.

\begin{itemize}
\item \textbf{Condition branch-driven prompt}: This prompt encourages the LLM to explore different control-flow paths, such as conditional statements and loops, improving branch and path coverage.
\item \textbf{Functionality-driven prompt}: This prompt focuses on the intended semantics of the method, helping the LLM generate test inputs that exercise typical behaviors and edge cases derived from parameter types and return values.
\item \textbf{Exception-oriented prompt}: This prompt drive the generation of test cases for robustness and failure handling, including invalid or unexpected inputs that should trigger exceptions or error states.
\end{itemize}

This multi-view prompt design is modular and extensible—new testing views (e.g., performance, concurrency) can be incorporated by adding corresponding prompt templates. By separating test planning from execution and grounding the design in testing theory, this step enhances the quality, coverage, and interpretability of generated tests.

\subsubsection{Test Method Transformation}
\label{sec:test-method-trans}

After designing structured test cases, \app transforms each test case group into an executable test method, aligning with how developers typically write tests—first decide what to test, then implement the logic.

\begin{figure}
    \centering
    \includegraphics[width=0.9\columnwidth]{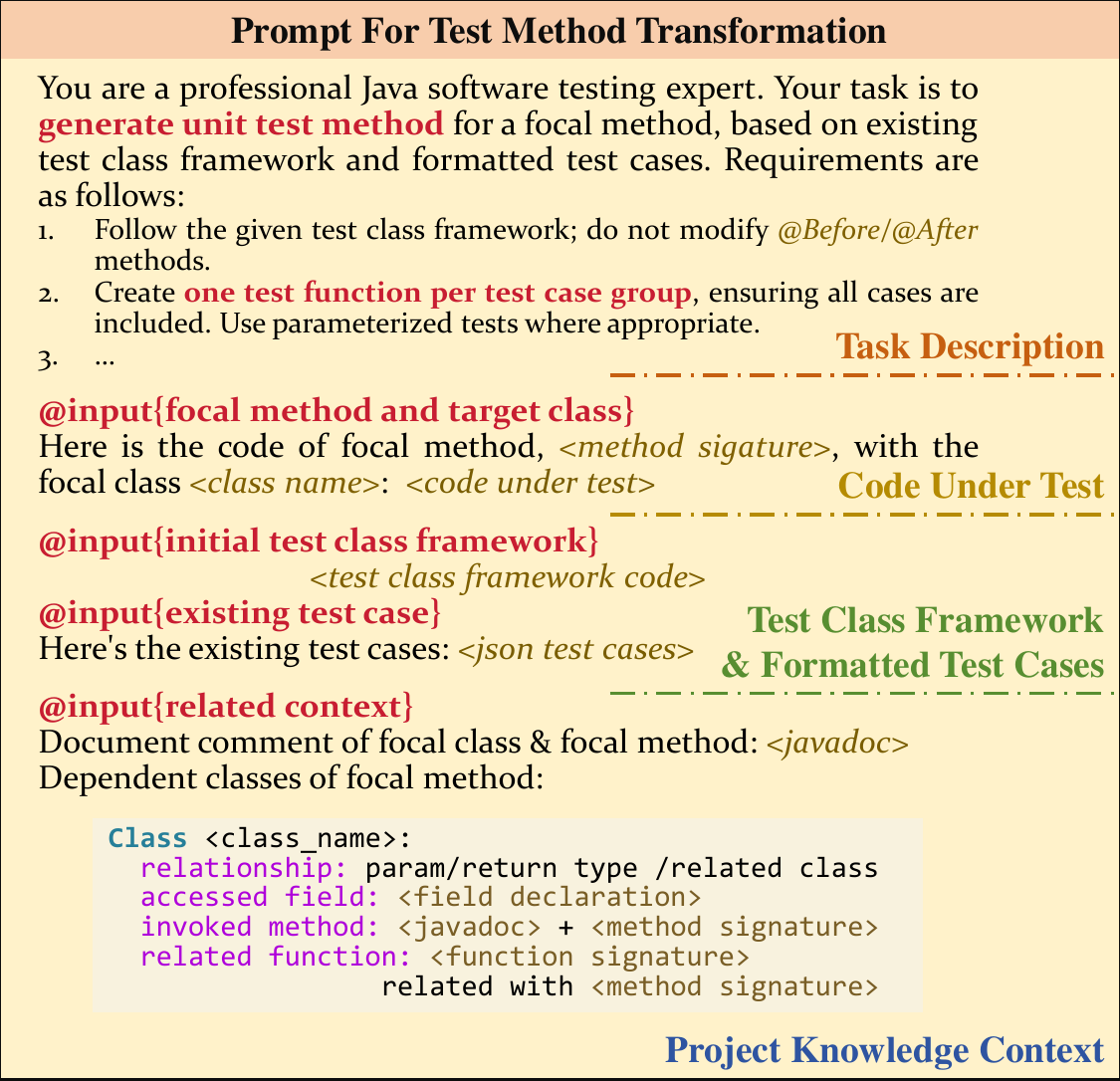}
    \caption{Prompt for test method trasformation.}
    \Description{Prompt for test method trasformation.}
    \label{fig:prompt-test-method-gen}
\end{figure}

\parabf{Prompt Construction.} We design a usage-aware prompt with several semantically distinct sections to guide the LLM in generating executable test methods, as shown in Figure~\ref{fig:prompt-test-method-gen}. First, test case groups contain structured cases sharing a common testing intent (e.g., valid inputs, edge cases, exceptions), each with a scenario, inputs, and expected output. The LLM generates one test method per group to ensure modularity and clarity. Second, the usage context, retrieved from the project knowledge base (Section~\ref{sec:usage_context}), provides focal method dependencies and related methods to guide realistic invocation and assertion generation. Lastly, the test class framework offers reusable setup/teardown code and field declarations, promoting consistency and eliminating redundancy.

\parabf{Usage Context Retrieval.}
To enhance generation quality, we retrieve methods that are functionally related to the focal method and its dependencies, offering a broader context of real-world usage patterns. Based on our observations, related methods are identified based on shared usage of functions and fields. We define the similarity between two functions 
$a$ and $b$ using a Jaccard-based metric~\cite{Jaccard1902} by combining with method usage similarity $Sim_{m}(a,b)$ and field usage similarity $Sim_{f}(a,b)$:

\begin{equation}
Sim(a,b)= Sim_{m}(a,b)+ Sim_f(a,b)
\end{equation}

\begin{equation}
Sim_m(a,b) = \frac{|M(a)\cap M(b)|}{|M(a)\cup M(b)|}
\end{equation}
\begin{equation}
Sim_f(a,b) = \frac{|F(a)\cap F(b)|}{|F(a)\cup F(b)|}
\end{equation}

Here, \(M(a)\) and \(F(a)\) denote the sets of methods and fields used by function \(a\), respectively. We select the top-N most similar functions to construct the context.

For each related class, we include: (1) the class declaration, optionally with Javadoc comments; (2) relevant field declarations accessed by dependent or related methods; (3) signatures of dependent methods with comments; and (4) signatures of related methods, annotated with markers indicating shared usage. This context provides the LLM with concise, relevant information to support realistic test generation.

\subsubsection{Test Class Integration}
After transforming individual test cases into test methods, the next step is to integrate them into a coherent and executable test class. This step ensures that all test methods, including newly generated ones and any accompanying setup or teardown logic, are correctly incorporated into the test class framework without introducing redundancy or conflicts.

To achieve this, \app performs static analysis on the generated methods to identify their roles and resolve duplicates. The integration process proceeds as follows:

\textbf{Setup/Teardown Merging.} If a newly generated method is identified (via static analysis) as a setup or teardown method (e.g., annotated with \texttt{@Before}, \texttt{@BeforeEach}, etc.), and an existing method of the same type already exists, we merge their bodies. Specifically, any code in the new method that is not present in the original is appended, ensuring initialization routines are preserved and extended without duplication.

\textbf{Test Method Deduplication.} For test methods and general helper methods, we identify duplicates by method name. When two methods share the same name, we further compare their bodies and retain the longer implementation as a proxy for logical richness. This deduplication addresses a common pattern where the LLM first generates simple placeholder tests and later generates more complete versions with the same or similar names, ensuring that the final output preserves the more comprehensive implementation.

\textbf{Non-conflicting Insertion.} For all other non-conflicting methods, we append them directly to the test class. This modular addition supports test diversity and extensibility without disrupting the original structure.
The integration process leverages standard AST-based static analysis tools to parse and manipulate the Java code structure safely and consistently. This automation ensures that the final test class is executable, maintainable, and free of duplicate or conflicting code.

By decoupling method generation from class integration, our approach maintains modularity while aligning with practical software engineering workflows, where tests are often extended iteratively and merged across sessions or contributors.

\subsubsection{Test Class Refinement}
In this stage, \app validates the generated test class by checking compilation and execution errors. Despite the rich contextual guidance, LLM outputs may still contain issues such as hallucinated import statements, improper mocking, incorrect usage of private methods, or runtime errors during test execution. To address these problems, we adopt an iterative refinement process with static validation and dynamic repair.

\parabf{Rule-based Repair.} Rule-based repair focuses on correcting straightforward and common errors detected by static analysis, such as hallucinated or missing import statements and unresolved symbols. Leveraging the previously constructed project-level test knowledge base, \app can automatically resolve dependencies by appending missing imports or correcting incorrect ones based on simple class names. The comprehensive indexing of classes, methods, and fields from the static analysis phase facilitates accurate dependency resolution without human intervention.

\parabf{LLM-based Repair.} For errors arising from semantic misunderstandings or subtle logic issues that cannot be fixed by static rules, \app applies LLM-based repair using compile-time errors, runtime logs, and failing test feedback. This information, combined with the focal method context and relevant function signatures, is incorporated into a repair prompt that is fed back to the LLM. The model then performs targeted correction of issues such as incorrect assertions, missing exception handling, or logic bugs causing runtime failures.

Importantly, this refinement process extends beyond compilation to include execution feedback. If runtime errors occur or tests fail due to assertion mismatches, these failure details are fed into the repair pipeline to enable iterative correction and improvement of the generated test code. This dynamic feedback loop helps ensure that the final test class is not only syntactically correct but also functionally robust and reliable.

\section{Evaluation}
\label{sec:evaluation}
In this section, we evaluate the effectiveness of \app by answering the following research questions (RQs):

\textbf{RQ1 (Effectiveness Comparison)}: Does \app outperform state-of-the-art baselines in terms of coverage scores and execution pass rate when testing complex methods?

\textbf{RQ2 (Ablation Study)}: How do the key components of \app contribute to its overall performance?

\textbf{RQ3 (User Study)}: Are the tests generated by \app more readable and maintainable than those produced by baseline methods?

\subsection{Experimental Setup}
\subsubsection{Dataset}
\label{sec:eval:dataset}
To evaluate the effectiveness of \app, we adopt the HITS dataset~\cite{wang_hits_2024}, which is constructed from 10 popular open-source Java projects spanning various domains (e.g., microservices, command-line tools, event engines). This dataset specifically targets complex methods—defined as those with cyclomatic complexity greater than 10—making it well-suited for assessing how effectively LLMs handle methods with intricate control flow and dependencies. The dataset comprises 110 tasks, each consisting of a focal method and its containing class.

\subsubsection{Baselines}
\label{sec:eval:baselines}
We compare \app with 4 state-of-the-art test generation methods: three are purely LLM-based, and one integrates search-based software testing (SBST) with LLMs. The LLM-based baslines adopt distinct strategies for context construction and test generation:

\textbf{ChatUnitTest}~\cite{chen_chatunitest_2024} provides the focal method and handcrafted context to the LLM, and applies iterative repair using error feedback from failed executions.

\textbf{ChatTester}~\cite{yuan2023no} is similar to ChatUnitTest but differs in how it constructs the focal method’s context, building it incrementally as needed.

\textbf{HITS}~\cite{wang_hits_2024} decomposes the focal method into slices and prompts the LLM to generate test cases for each slice, which are then merged into a complete test suite. This fine-grained approach helps mitigate the difficulty LLMs face when reasoning about complex logic holistically.

For fair comparison, we adapt official or open-source implementations of these methods~\cite{zju-chatunitest_2025}. ChatUnitTest and ChatTester generate one test class per focal method, while HITS’s slice-level tests are merged into a single class to ensure consistent execution and setup. All baselines and \app use the same LLM backend—\texttt{gpt-4o-mini}~\cite{chatgpt_model}—with temperature=0.5, ensuring consistency in generation quality, efficiency, and cost.

The SBST–LLM hybrid method, \textbf{UTGen}~\cite{deljouyi2025leveraging}, employs a traditional SBST tool to generate initial test cases, then leverages an LLM to rename functions and variables in order to improve test readability. Since UTGen rely on an LLM's ability on code understanding instead of enhancing code coverage, we follow the initial configuration provided in its replication package~\cite{UTGen} (EvoSuite~\cite{fraser2011evosuite} as the generation tool and CodeLlama-7b~\cite{codellama} used for refinement) when conducting our experiments to evaluate its performance.

\begin{table}
\centering
\caption{Detailed Information of HITS Dataset}
\label{tab:dataset}
\resizebox{\columnwidth}{!}{%
\begin{tabular}{llll}
\hline
Project                  & Domain               & Version        & \#MUT \\ \hline
Commons-CLI              & Cmd-line Interface   & 1.7.0-SNAPSHOT & 2                  \\
Commons-CSV              & Data processing      & 1.10.0         & 6                  \\
Gson                     & Serilization         & 2.10.1         & 20                 \\
Commons-codec            & Encoding             & 3a6873e        & 18                 \\
Commons-collections4     & Utility              & 4.5.0-M1       & 14                 \\
JDom2                    & Text Processing(XML) & 2.0.6          & 21                 \\
Datafaker                & Data Generation      & 1.9.0          & 6                  \\
Event-ruler              & Event Engine         & 1.4.0          & 15                 \\
windward                 & Micoservices         & 1.5.1-SNAPSHOT & 2                  \\
batch-processing-gateway & Cloud Computing      & 1.1            & 6                  \\ \hline
\end{tabular}%
}
\end{table}

\subsection{RQ1: Effectiveness Comparison}
\label{sec:rq1}
\subsubsection{Design}
We use the HITS dataset~(Section~\ref{sec:eval:dataset}), which contains 110 test generation tasks from 10 open-source Java projects. Our method builds the project knowledge base by analyzing the corresponding project source code versions in the dataset. For these tasks, both \app and baseline methods generate unit tests using GPT-4o-mini as the backbone model. 
For fair comparison, KTester and all baselines generate exactly one test class per focal method, with no restriction on the number of test cases or generation time within each class. This allows complex methods to achieve high coverage through diverse test cases. We limited automatic repair iterations to 5 in \app. To mitigate the effect of randomness introduced by LLMs, we repeated each experiment five times and report the average results.

To systematically evaluate the effectiveness of generated unit tests, we adopt eight widely used metrics for unit test quality, capturing correctness, sufficiency and efficiency ~\cite{yuan2023no,wang_hits_2024,chen_chatunitest_2024,gu_testart_2025}.

\begin{itemize}
\item \textbf{Compile Pass Rate (CPR)}: percentage of test classes that compile successfully.
\item \textbf{Execution Pass Rate (EPR)}: percentage that run without runtime errors or assertion failures.
\item \textbf{Line Coverage (LC)}: ratio of executable lines in the focal method covered by all generated tests.
\item \textbf{Branch Coverage (BC)}: percentage of conditional branches covered.
\item \textbf{Line Coverage of Passed Tests (LCP)}: line coverage computed only on tests that compile and execute successfully.
\item \textbf{Branch Coverage of Passed Tests (BCP)}: branch coverage computed similarly on passing tests.
\item \textbf{Average Time pre Task (AvT)}: average time required by a method to generate test class for a single task.
\item \textbf{Average Test Cases Number (AvTC)}: average number of test cases contained in each generated test class.
\end{itemize}

Coverage is measured using Jacoco~\cite{Jacoco},  an open-source tool for measuring code coverage in Java projects, ensuring objective and consistent evaluation. By combining correctness (CPR, EPR), coverage metrics (LC, BC, LCP, BCP) and efficiency metrics (AvT, AvTC), we provide a balanced and robust assessment of test quality, facilitating fair comparison between \app and baseline approaches.

\subsubsection{Results}

Table~\ref{tab:RQ1-result} presents the effectiveness comparison across four methods. Overall, \app achieves the best performance on all six evaluation metrics, demonstrating its ability to generate high-quality unit tests both in terms of correctness and coverage.

\parabf{Correctness.}
\app reaches a perfect Compile Pass Rate (CPR) of \textbf{100\%}, indicating that all generated test classes are syntactically valid and type-correct. 
While UTGen also achieves 100\% CPR, the other LLM-based baselines fall short, suggesting potential issues in code generation or dependency handling. Regarding the Execution Pass Rate (EPR), which measures runtime correctness, \app attains 77.07\%, slightly lower than UTGen. This is expected, as UTGen builds upon EvoSuite-generated tests, which provide a strong foundation for runtime correctness. Nevertheless, \app still surpasses all other LLM-based methods in EPR, demonstrating its effectiveness in generating executable and correct test cases. This result indicates that incorporating project knowledge yields substantial benefits, as it more effectively guides LLMs toward generating executable test code.

\parabf{Sufficiency.}
In terms of code coverage, \app achieves the highest Line Coverage (LC) of \textbf{63.94\%} and Branch Coverage (BC) of \textbf{55.46\%}, significantly outperforming HITS (52.27\% LC, 45.93\% BC) while generating less test case (7.33 vs. 15.78). ChatTester yields the lowest coverage, highlighting its limited ability to explore program logic.
UTGen is constrained by EvoSuite’s inability to operate on certain project-specific packages (e.g., \textit{com.apple.spark} in the batch-processing-gateway project) and by the path-explosion problem inherent to SBST techniques. As a result, its coverage exceeds only that of ChatTester.
When restricting evaluation to only the tests that both compile and execute successfully, \app maintains the lead with \textbf{55.52\%} Line Coverage of Passed Tests (LCP) and \textbf{47.21\%} Branch Coverage of Passed Tests (BCP), confirming the practical adequacy of its outputs.

% Please add the following required packages to your document preamble:
% \usepackage{graphicx}
\begin{table}[]
\centering
\caption{Comparison of Effectiveness across Baselines}
\label{tab:RQ1-result}
\resizebox{\columnwidth}{!}{%
\begin{tabular}{llllll}
\hline
    & UTGen          & ChatTester & ChatUniTest & HITS          & \app \\ \hline
CPR & \textbf{100}   & 55.18      & 98.97       & 97.82         & \textbf{100}        \\
EPR & \textbf{90.05} & 50.17      & 65.26       & 71.38         & \underline{76.41}       \\
LC  & 31.17          & 28.51      & 44.64       & \underline{52.27} & \textbf{63.94}      \\
BC  & 28.72          & 24.94      & 38.06       & \underline{45.93} & \textbf{55.46}      \\
LCP & 30.69          & 22.26      & 33.47       & \underline{43.85} & \textbf{55.52}      \\
BCP & 28.11          & 19.63      & 28.51       & \underline{38.81} & \textbf{47.21}      \\ 
AvT (s) & 1068  & 354.83     & \underline{200.96}      & 625.69  & \textbf{152.68}        \\
AvTC & 9.23 & 3.32 & 4.77 & 15.78 & 7.33 \\ \hline
\end{tabular}%
}
\end{table}

\begin{table}[]
\caption{Comparison of Effectiveness across KTester implemented with different models}
\label{tab:disscuss-model}
\resizebox{0.9\columnwidth}{!}{%
\begin{tabular}{l|ccc}
\hline
    & KTester-gpt & KTester-claude & KTester-deepseek \\ \hline
CPR & 100         & 100            & 93.92            \\
EPR & 76.41       & 81.75          & 82.41            \\
LC  & 63.94       & 66.46          & 71.22            \\
BC  & 55.46       & 60.66          & 65.75            \\
LCP & 55.52       & 56.30          & 65.19            \\
BCP & 47.21       & 50.76          & 59.33            \\ \hline
\end{tabular}%
}
\end{table}

\begin{figure}[thbp]
    \centering
    \includegraphics[width=0.9\columnwidth]{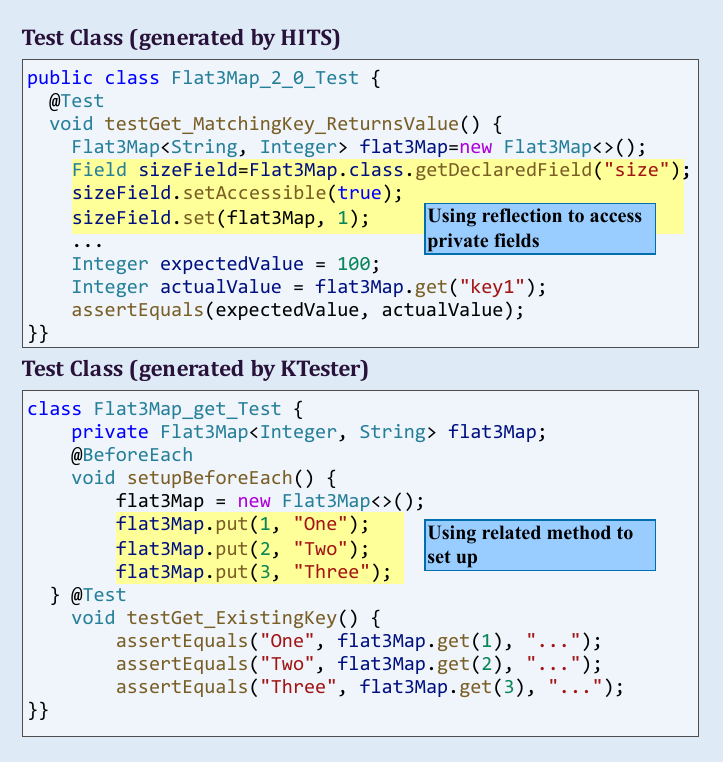}
    \caption{Test Classes generated by HITS and \app.}
    \Description{Test Classes generated by HITS and \app.}
    \label{fig:rq1:example}
\end{figure}

These results provide evidence that integrating project-specific and test-domain knowledge into LLM-based generation pipelines is associated with improved test adequacy and correctness, as reflected by higher coverage and pass rates. In particular, \app demonstrates stronger capability in exercising program logic and control flow, making it a more effective and practical solution for automated unit test generation.

Figure \ref{fig:rq1:example} shows the test class generated for the \textit{get()} method with HITS and \app. HITS used  reflection to access and set private values, which increases the difficulty of understanding. In fact, since Flat3Map is a Map data structure, we can better modify private fields through the relevant methods \textit{put ()} in this class, thus covering every conditional branch during testing.

\parabf{Generalizability.}
To assess whether KTester maintains its performance when implemented with alternative LLMs, we implemented KTester using claude-3.5-haiku-20241022~\cite{claude-35-haiku} and deepseek-v3.1~\cite{deepseek-v31}, and the experimental results are reported in Table~\ref{tab:disscuss-model}, which achieved even better results than gpt4o-mini, confirming its generalizability.

\finding{\app outperforms all baselines across eight metrics. Compared to the strongest baseline (HITS), and while generating on average 8.45 fewer test cases, it improves execution pass rate by 5.03\% and line coverage by 11.67\%, demonstrating clear gains in both correctness and sufficiency.}

\subsection{RQ2: Ablation Study}
\label{sec:rq2}
\subsubsection{Design}
To assess the individual impact of key components in \app on unit test generation, we conducted an ablation study. Four variants were created, each removing or modifying one component while keeping the rest of the pipeline intact. This design isolates each component’s effect on correctness and coverage. The variants include:

\begin{itemize}
    \item \textbf{\app-UTE}: without the procedure of \textbf{u}sage \textbf{t}race \textbf{e}xtraction (section~\ref{sec:knowledge_extraction}).
    \item \textbf{\app-FMR}: without the step of \textbf{f}unctionally related \textbf{m}ethod \textbf{r}etrieval (section~\ref{sec:usage_context}).
    \item \textbf{\app-MVG}: using only conditional branch guidance instead of the whole \textbf{m}ulti-\textbf{v}iew guidance strategy (section~\ref{sec:test-case-design}).
    \item \textbf{\app-DGT}: replace the processes of test cases design and test method transformation (section~\ref{sec:test-method-trans}) with a single process that \textbf{d}irectly \textbf{g}enerates \textbf{t}est methods using a multi-view guidance strategy.
\end{itemize}

Using the same experimental setup and metrics as in RQ1, we evaluated all variants and compared their results to the original \app to measure each component’s contribution to test quality.

\subsubsection{Results}

Table~\ref{tab:RQ2-result} summarizes the performance of \app variants. All variants exhibit performance drops across correctness and coverage metrics, confirming that each component contributes to \app. Notably, \app-UTE shows the largest correctness degradation, with CPR and EPR decreasing by 6.37\% and 14.41\%, respectively, highlighting the critical role of usage trace extraction in improving test correctness. \app-DGT leads to the most substantial coverage losses, with LC and BC dropping by 13.39\% and 11.15\%, respectively, indicating that separating test case design from test method generation is crucial for test adequacy. The other variants—\app-FMR and \app-MVG have comparatively smaller impacts across both correctness and coverage metrics, but also result in performance reductions.

\finding{Usage trace extraction (\app-UTE) is the primary contributor to test correctness, while modular test case transformation (\app-DGT) contributes most to test adequacy. Other components also contribute positively with smaller and less distinguishable effects.}

% Please add the following required packages to your document preamble:
% \usepackage{graphicx}

\begin{table}[]
\centering
\caption{Comparison of Effectiveness across Variants and original \app.}
\label{tab:RQ2-result}
\resizebox{\columnwidth}{!}{%
\begin{tabular}{l|llllll}
\hline
    & CPR   & EPR   & LC    & BC    & LCP   & BCP   \\ \hline
\app-UTE & 93.63   & 62.00 & 57.53 & 51.27 & 45.37 & 39.83 \\
\app-FMR & 95.66 & 66.99 & 59.17 & 52.32 & 46.83 & 41.02 \\
\app-MVG & 97.45   & 75.31 & 53.67 & 46.26 & 45.48 & 39.36 \\
\app-DGT & 96.14 & 64.96 & 50.55 & 44.31 & 41.07 & 35.65 \\ \hline
\app & \textbf{100} & \textbf{76.41} & \textbf{63.94} & \textbf{55.46} & \textbf{55.52} & \textbf{47.21} \\ \hline
\end{tabular}%
}
\end{table}

\subsection{RQ3: User Study}
\label{sec:rq3}
\subsubsection{Design}
\looseness=-1
To investigate whether the tests generated by our approach exhibit superior readability and maintainability, we conducted a user study targeting professional developers. We first selected 10 tasks from the dataset where both \app and the baseline methods successfully produced compilable test cases. Ensuring compilability allows participants to concentrate on qualitative properties of the test code—such as clarity, structure, and maintainability—without being influenced by technical correctness issues.

We recruited participants through a public invitation distributed across the computer science departments of six universities. From the volunteers, we selected 15 participants (5 Ph.D. students and 10 Master’s students), all with 2–5 years of Java development experience. Participants received compensation to encourage careful and unbiased evaluation.

For each task, participants were presented with the focal method under test, its API documentation and relevant class context, then the related test classes generated by different methods. To avoid bias, the identities of the methods were fully anonymised, and the presentation order was randomised.
Participants were instructed to respond to the following three questions, each designed to target a key quality dimension of the generated test code:

\begin{enumerate}
    \item To what extent do you agree that the unit test class is functionally correct? Specifically, does it adequately cover relevant input combinations, contain at least one meaningful assertion per test method, and reliably detect functional faults through deterministic and repeatable execution?
    \item To what extent do you agree that the unit test class is highly readable? For instance, are the method and variable names clear and descriptive? Does each test follow the Arrange–Act–Assert structure to clearly communicate intent? Are assertion failure messages informative and helpful for debugging?
    \item To what extent do you agree that the unit test class is maintainable? For example, does it minimise redundancy or duplicated code? Is the test class support the easy addition of new test cases? Can developers easily update affected tests when the source code changes?
\end{enumerate}

Each question is rated on a four-point Likert scale—strongly agree, somewhat agree, somewhat disagree, and strongly disagree. This setup enables quantitative comparison of subjective test quality across approaches. Following prior work \cite{yuan2023no, icse25PanKKPS25}, we adopt an even-numbered scale to avoid ambiguous “neutral” responses, and clearly separates positive from negative judgments.

\begin{figure}
    \centering
    \includegraphics[width=\columnwidth]{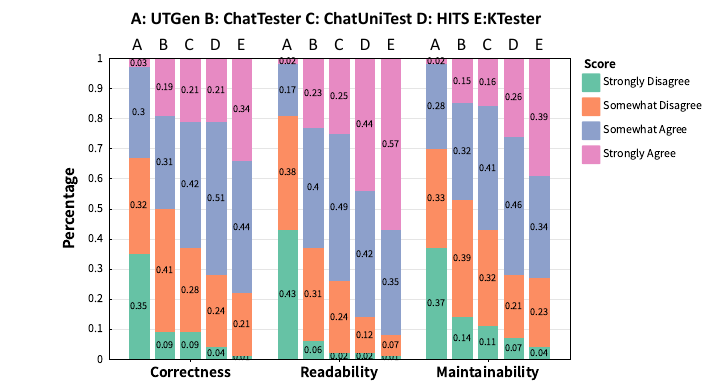}
    \caption{Score Distribution of Correctness, Readability and Maintainability}
    \Description{Score Distribution of Correctness, Readability and Maintainability}
    \label{fig:RQ3-result}
\end{figure}

\subsubsection{Results}

Figure~\ref{fig:RQ3-result} presents the distribution of developer ratings across the three evaluation dimensions. Across all three dimensions, our method (\app, labeled as E) consistently outperforms the baselines in terms of perceived quality.

In the dimension of Correctness, \app achieves the highest proportion of “Strong Agree” ratings (0.34), while keeping “strongly disagree” scores low (0.01). This trend indicates a strong alignment with expected test behavior as perceived by developers. In terms of Readability, \app demonstrates the most favorable distribution, with over 90\% of responses rated as “strongly Agree” (0.57) or “Somewhat Agree” (0.35), suggesting that participants found its tests clearer and more accessible compared to those from other methods. For Maintainability, \app has the highest proportion of "Strong Agree" responses (0.39) and the lowest proportion of “strongly disagree” scores (0.04). This suggests that developers found the structure and logic of the tests generated by our approach easier to understand and adapt.

Interestingly, although UTGen uses an LLM to improve test readability, it still received more than 67\% negative responses across all dimensions. This is mainly due to (1) function signatures and variable names often remaining unmodified, leaving the test code hard to understand, and (2) UTGen’s EvoSuite component testing private methods via public callers, while LLM-based methods rely on reflection, making participants feel hard to identify impacted tests after source code changes and thus negatively affecting assessments of correctness and maintainability.

\finding{\app outperforms all baselines in human evaluation across correctness, readability, and maintainability.}

\section{Related Work}
In this section, we mainly introduce related work on LLM-based unit test generation.

\subsection{LLM-based Unit Test Generation}
Large language models (LLMs) have become a powerful tool for automated unit test generation, offering human-like test code that overcomes the readability and maintainability limitations of traditional methods~\cite{harman2015achievements, mcminn2011search, baldoni2018survey, cadar2008klee, cha2012unleashing}. Existing LLM-based approaches can be broadly categorized into two paradigms: fine-tuning and prompt-based methods.

\textbf{Fine-tuning} approaches frame test generation as a supervised sequence-to-sequence task, learning mappings from focal methods to complete test cases using curated corpora, \eg AthenaTest~\cite{tufano2020unit}, TeCo~\cite{nie2023learning}, and subsequent variants~\cite{dakhel2024effective,deng2024large,deng2023large}, which explore different encoder–decoder architectures to capture the semantic relationships between code and tests. 
EXLONG~\cite{zhang2024exlong} further fine-tunes CodeLlama for exception-oriented test generation. Unlike these approaches, our method injects project and testing knowledge through prompt design and generation workflows, enabling greater flexibility and compatibility with future LLMs.

\textbf{Prompt-based} methods, in contrast, exploit the zero-shot reasoning abilities of instruction-tuned models. Systems such as ChatTester~\cite{yuan2023no}, ChatUniTest~\cite{chen_chatunitest_2024}, and ASTER~\cite{icse25PanKKPS25} use carefully designed prompts and multi-step reasoning to guide models like GPT in test identification, input generation, and assertion construction. Extensions incorporate additional context or search strategies, including documentation and usage examples in TestPilot~\cite{schafer2023empirical} and IDE-based search in TestSpark~\cite{sapozhnikov2024testspark}.

Recent work explores hybrid approaches that combine LLMs with complementary techniques. CODAMOSA~\cite{lemieux2023codamosa} integrates LLM-generated test seeds with evolutionary algorithms to improve coverage-driven test generation, CoverUp~\cite{altmayer2025coverup} refines tests through coverage-guided dialogues, and SymPrompt~\cite{ryan2024code} leverages symbolic execution to steer prompting.

Despite these advancements, most approaches lack project-specific context (\eg instantiation patterns, dependencies, and API usage) and overlook core testing principles such as boundary analysis, equivalence partitioning, and exception handling, yielding tests without execution error but insufficiently validate functionality.

\subsection{Knowledge-enhanced LLMs for SE Tasks}
Recent studies have explored enhancing LLM performance on SE tasks by incorporating task-relevant context. In APR, Xia et al.~\cite{Xia2023} show that simply providing the buggy function can outperform traditional tools, while ChatRepair~\cite{XiaAuto2024} leveraged failing test names and assertions for interactive prompting. Other work enriches context with bug-localized code~\cite{Prenner2024}, relevant identifiers~\cite{XiaPlastic2024,JiangImpact2023}, stack traces~\cite{haque-etal-2025-towards}, and bug reports~\cite{FakhouryNL2Fix2024}.

In parallel, retrieval-augmented generation (RAG) integrates external knowledge by retrieving relevant information from codebases or databases. For code completion, ReACC~\cite{lu2022reaccretrievalaugmentedcodecompletion} retrieves semantically similar code snippets, and Wu et al.~\cite{WuREPOFORMER2024} propose a selective RAG framework to reduce redundancy. For code generation, recent work~\cite{li2025coderagsupportivecoderetrieval,Athale_2025} represents code repositories as graphs or structured knowledge to retrieve and add repository-level context to LLMs.
\section{Threats to Validity}
\label{sec:threats}

A key threat lies in the use of LLMs. To ensure fairness, all approaches are evaluated using the same model version. For baselines, we rely on official implementations or carefully reimplement them following published details, verifying outputs to align with reported results. Benchmark tasks and ground-truth tests are directly reused from prior work for consistency.

Another threat concerns generality and subjectivity. Although experiments focus on Java for comparability~\cite{yuan2023no,wang_hits_2024}, \app{} is not inherently tied to Java. Extending to other frameworks mainly requires prompt and library updates, while supporting other languages only involves replacing the AST analysis tool (e.g., tree-sitter), resulting in low adaptation cost. Future work will explore such extensions. While correctness and coverage are objective metrics, they may not fully capture clarity and structure. we mitigate this via a user study on readability and maintainability with 14 professional developers, though results may still reflect participant bias and limited sampling. We believe this is sufficient, but a broader sampling could further strengthen validity.
\section{Conclusions}
We present \app, a project-aware, testing-domain-knowledge-guided framework for LLM-based unit test generation. By extracting comprehensive project knowledge and decoupling test case design from test code generation, \app effectively guides the LLM to produce semantically rich and maintainable tests. Our multi-perspective prompting strategy ensures diverse and thorough coverage of testing scenarios, while structured test templates enhance code clarity and reusability. Extensive evaluations on real-world open-source projects show that \app consistently outperforms existing state-of-the-art methods in correctness, coverage, readability, and maintainability. Future work explores integrating dynamic analysis for enriching project knowledge and extending the approach to other languages and testing paradigms.

\begin{acks}
This work was supported by National Natural Science Foundation of China (Grant No. 62402113), the General Program of the Natural Science Foundation of Guangdong Province, China (Grant No. 2025A1515011631), Social Science Planning Project (Grant No. SZ2025A002), and GMCC-SYSU Joint Lab for Smart Applications.
\end{acks}

% \normalem
\bibliographystyle{ACM-Reference-Format}
% \balance
\bibliography{refs}

\end{document}